\documentclass[prd,aps,showpacs,nofootinbib,tightenlines]{revtex4}
\usepackage {amsmath}
\usepackage {amssymb}
\usepackage {epsfig}
\usepackage {graphicx}
\usepackage {booktabs}

\begin{document}
\title{Quasi-two-body decays $B_{(s)}\to P\rho^\prime(1450), P\rho^{\prime\prime}(1700)\to P\pi\pi$ in the
perturbative QCD approach}
\author{Ya Li$^{2}$}               \email{liyakelly@163.com}
\author{Ai-Jun Ma$^2$}         \email{theoma@163.com}
\author{Wen-Fei Wang$^1$}       \email{wfwang@sxu.edu.cn}
\author{Zhen-Jun Xiao$^{2,3}$}  \email{xiaozhenjun@njnu.edu.cn}
\affiliation{$^1$ Institute of Theoretical Physics, Shanxi University, Taiyuan, Shanxi 030006, China}
\affiliation{$^2$ Department of Physics and Institute of Theoretical Physics,
                          Nanjing Normal University, Nanjing, Jiangsu 210023, People's Republic of  China}
\affiliation{$^3$ Jiangsu Key Laboratory for Numerical Simulation of Large Scale Complex
Systems, Nanjing Normal University, Nanjing, Jiangsu 210023, People's Republic of China}
\date{\today}

\begin{abstract}
In this work, we calculate the $CP$-averaged branching ratios and direct $CP$-violating asymmetries of the
quasi-two-body decays $B_{(s)}\to P \rho^\prime(1450), P\rho^{\prime\prime}(1700)\to P \pi\pi$
by employing the perturbative QCD (PQCD) factorization approach, where $P$ is a light pseudoscalar
meson $K, \pi, \eta$, and $\eta^{\prime}$.
The considered decay modes are studied in the quasi-two-body framework by parametrizing the
two-pion  distribution amplitude $\Phi_{\pi\pi}^{\rm P}$.
The $P$-wave timelike form factor $F_{\pi}$ in the resonant regions
associated with the $\rho^\prime(1450)$ and $\rho^{\prime\prime}(1700)$ is estimated based on available
experimental data.
The PQCD predictions for the $CP$-averaged branching ratios of the decays
$B_{(s)}\to P\rho^\prime(1450), P\rho^{\prime\prime}(1700)\to P\pi\pi$ are in the order of $10^{-7}- 10^{-9}$.
The branching ratios of the two-body decays $B_{(s)}\to P\rho^\prime(1450), P\rho^{\prime\prime}(1700)$ are
extracted from the corresponding quasi-two-body decay modes.
The whole pattern of the squared pion form factor  $|F_\pi|^2$ measured by {\it BABAR} Collaboration could also be
understood based on our studies.
The PQCD predictions in this work will be tested by the precise data from the LHCb and the future Belle II experiments.
\end{abstract}


\maketitle


\section{Introduction}
In recent years,  prompted by a large number of experimental measurements~\cite{bfbook,prd78-072006,prd79-072004,lhcb0,prl111-101801,jhep10-143,prd90-112004,prl112-011801,prd95-012006,hfag2016},
three-body hadronic $B$-meson decays have been studied by using different theoretical frameworks~\cite{plb564-90,prd91-014029,CY16,Wang-2014a,Wang-2015a,Wang-2016,ly15,ly16,zhou17}.
For such three-body decays, both resonant and nonresonant contributions may appear, as well as the possible final state
interactions~\cite{prd89-094013,1512-09284,89-053015}. The nonresonant contributions have been studied with the method of heavy meson chiral perturbation theory~\cite{prd46-1148,prd45-2188,plb280-287} in Ref.~\cite{CY16}.
Meanwhile, the resonant contributions are usually described with the isobar model~\cite{prd11-3165}
in terms of the Breit-Wigner formalism~\cite{BW-model}.
Based on the QCD-improved factorization~\cite{prl83}, such decays have been studied by many authors
~\cite{plb622-207,prd74-114009,B.E:2009th,ST:15,CY01,prd76-094006,CY16,prd89-094007,prd87-076007}.
By employing the perturbative QCD (PQCD) approach, the $B \to 3h$ decays have also been investigated in
Refs.~\cite{Wang-2014a,Wang-2015a,Wang-2016,ly15,ly16,zhou17,ma16,ma17,Chen:2002th,chen:2004th}.

In the PQCD approach~\cite{Chen:2002th,chen:2004th} for the cases of a $B$-meson decaying into three final states,
we restrict ourselves to the specific kinematical configurations, in which two energetic mesons are almost collimating to
each other.
The contribution from the region, where there is at least one pair of light mesons having an invariant mass below
$O(\bar\Lambda m_B)$, 
$\bar\Lambda=m_B-m_b$ being the $B$ meson and $b$ quark mass difference,
as discussed in Refs.~\cite{Wang-2014a,Wang-2015a,Chen:2002th,chen:2004th}, is assumed dominant.
The final state interactions are expected to be suppressed in such conditions.
As a result, the dynamics associated with the meson pair could be factorized into a two-meson distribution amplitude
$\Phi_{h_1h_2}$~\cite{MP,MT01,MT02,MT03,MN,Grozin01,Grozin02}.
The typical PQCD factorization formula for the $B\to h_1h_2h_3$ decay amplitude can be written in the form of ~\cite{Chen:2002th}
\begin{eqnarray}
\mathcal{A}=\Phi_B\otimes H\otimes \Phi_{h_1h_2}\otimes\Phi_{h_3}.
\end{eqnarray}
The hard kernel $H$ describes the dynamics of the strong and electroweak interactions in the three-body
hadronic decays in a similar way as the cases of the two-body $B\to h_1 h_2$ decays,
and $\Phi_B$ and $\Phi_{h_3}$ are the wave functions for the $B$ meson
and the final state $h_3$, which absorb the nonperturbative dynamics in the related processes.

In this work, we extend the previous studies~\cite{Wang-2016,ly16} to the decays $B \to P\rho^{\prime}(1450) \to P\pi\pi$ and
$B \to P\rho^{\prime\prime}(1700) \to P\pi\pi$ in the PQCD approach with the help of the two-pion distribution
amplitudes $\Phi_{\pi\pi}^{\rm P}$, where the $P$ stands for the light pseudoscalar mesons, $P=(\pi, K, \eta$, or $\eta^\prime)$.
For simplicity, in the following parts of this work, $\rho^\prime$ and $\rho^{\prime\prime}$ will be adopted
to take the place of $\rho^{\prime}(1450)$ and $\rho^{\prime\prime}(1700)$, respectively.
The theoretical studies of the excited states will provide us with a deeper understanding of the internal structure of hadrons.
For $\rho^\prime$ and $\rho^{\prime\prime}$, there are
not many studies except Refs.~\cite{prd60-094020,prc79-025201,prd77-116009,1205-6793} in the frameworks of the quark model,
the large-$N_c$ limits,  or the double-pole QCD sum rules.
For the phenomenological study of the two-body decays $B \to P\rho^{\prime}$ and $B \to P\rho^{\prime\prime}$,
we still lack the distribution amplitudes of the states $\rho^{\prime}$ and $\rho^{\prime\prime}$ at present.
Fortunately, we are allowed to single out the $\rho^{\prime}$(and $\rho^{\prime\prime}$)
component according to the two-pion distribution amplitudes
$\Phi_{\pi\pi}^{\rm P}$ as has been done in Ref.~\cite{Wang-2016}. Following Ref.~\cite{Wang-2016},
we here make an attempt to study the $B \to P\rho^{\prime} \to P \pi\pi$ and $B \to P\rho^{\prime\prime} \to P \pi\pi$ decays in
the quasi-two-body framework based on the PQCD factorization approach.
And the branching fractions for the two-body decays $B \to P \rho^{\prime} (\rho^{\prime\prime})$ will be extracted
from the quasi-two-body processes $B \to P \rho^{\prime} (\rho^{\prime\prime}) \to P \pi\pi$.

This paper is organized as follows. In Sec.~II, we give a brief introduction for the theoretical framework.
The numerical values, some discussions, and the conclusions will be given in the last two sections.


\section{FRAMEWORK}\label{sec:2}  
In the light-cone coordinates,
the $B$ meson momentum $p_{B}$, the total momentum of the pion pair,
$p=p_1+p_2$, the momentum $p_3$ of the final state meson $P$, the momentum $k_B$ of the spectator quark in the $B$ meson,
the momentum $k$ for the resonant state $\rho^{\prime}(\rho^{\prime\prime})$, and $k_3$ for the final state $P$ are chosen as
\begin{eqnarray}\label{mom-pBpp3}
p_{B}&=&\frac{m_{B}}{\sqrt2}(1,1,0_{\rm T}),~\quad p=\frac{m_{B}}{\sqrt2}(1,\eta,0_{\rm T}),~\quad
p_3=\frac{m_{B}}{\sqrt2}(0,1-\eta,0_{\rm T}),\nonumber\\
k_{B}&=&\left(0,x_B \frac{m_{B}}{\sqrt2} ,k_{B \rm T}\right),\quad k= \left( z\frac{m_{B}}{\sqrt2},0,k_{\rm T}\right),\quad
k_3=\left(0,(1-\eta)x_3 \frac{m_B}{\sqrt{2}},k_{3{\rm T}}\right), \label{mom-B-k}
\end{eqnarray}
where $m_{B}$ is the mass of $B$ meson, and the variable $\eta$ is defined as $\eta=\omega^2/m^2_{B}$ with
the invariant mass squared $\omega^2=p^2=(p_1+p_2)^2$.
The parameter $x_B, z, x_3$ denotes the momentum fraction of the positive quark in each meson and runs from zero to unity.
$k_{B \rm T}, k_{\rm T}$, and $k_{3{\rm T}}$ denote the transverse momentum of the positive quark, respectively.
If we choose $\zeta=p^+_1/p^+$ as one of the pion pair's momentum fractions, the two pions momenta
$p_{1,2}$ can be written as
\begin{eqnarray}
 p_1=(\zeta \frac{m_B}{\sqrt{2}}, (1-\zeta)\eta \frac{m_B}{\sqrt{2}}, p_{1\rm T}),~\quad
 p_2=((1-\zeta)\frac{m_B}{\sqrt{2}}, \zeta\eta \frac{m_B}{\sqrt{2}}, p_{2\rm T}).
\end{eqnarray}

The two-pion distribution amplitudes can be described in the same way as in Ref.~\cite{Wang-2016} ,
\begin{eqnarray}
\Phi_{\pi\pi}^{\rm P}=\frac{1}{\sqrt{2N_c}}[{ p \hspace{-2.0truemm}/ }\Phi_{v\nu=-}^{I=1}(z,\zeta,\omega^2)+\omega\Phi_{s}^{I=1}(z,\zeta,\omega^2)
+\frac{{p\hspace{-1.5truemm}/}_1{p\hspace{-1.5truemm}/}_2
  -{p\hspace{-1.5truemm}/}_2{p\hspace{-1.5truemm}/}_1}{w(2\zeta-1)}\Phi_{t\nu=+}^{I=1}(z,\zeta,\omega^2)] \;,
\label{eq:phifunc}
\end{eqnarray}
with
\begin{eqnarray}
\Phi_{v\nu=-}^{I=1}&=&\phi_0=\frac{3F_{\pi}(s)}{\sqrt{2N_c}}z(1-z)\left[1
+a^0_2\; \frac{3}{2}\left [ 5(1-2z)^2-1 \right ]\right] P_1(2\zeta-1) \;, \label{eq:phi1}\\
\Phi_{s}^{I=1}&=&\phi_s=\frac{3F_s(s)}{2\sqrt{2N_c}}(1-2z)\left[1
+a^s_2\; \left ( 10z^2-10z+1 \right )\right] P_1(2\zeta-1) \;, \label{eq:phi2}\\
\Phi_{t\nu=+}^{I=1}&=&\phi_t=\frac{3F_t(s)}{2\sqrt{2N_c}}(1-2z)^2\left[1+a^t_2\; \frac{3}{2}\;
\left [ 5(1-2z)^2-1 \right]\right] P_1(2\zeta-1) \;,
\label{eq:phi3}
\end{eqnarray}
where the Legendre polynomial $P_1(2\zeta-1)=2\zeta-1$.

After taking the $\rho$-$\omega$ interference and excited-state contributions into account,
the timelike form factor $F_\pi(s)$ in Eq.~(\ref{eq:phi1}) can be written in the following form~\cite{prd86-032013}:
\begin{eqnarray}
F_{\pi}(s)&=& \left[ {\rm GS}_{\rho}(s,m_{\rho},\Gamma_{\rho})
\frac{1+c_{\omega} {\rm BW}_{\omega}(s,m_{\omega},\Gamma_{\omega})}{1+c_{\omega}}+\sum_i\; c_i\; {\rm GS}_i(s,m_i,\Gamma_i) \right]
\left[ 1+\sum_i c_i\right] ^{-1}\;, \label{eq:fpp}
\end{eqnarray}
where $s=\omega^2=m^2(\pi\pi)$ is the two-pion invariant mass square, and $\Gamma_i$ ($m_i$) is the decay
width (mass) for the relevant resonance $i=(\rho^{\prime}, \rho^{\prime \prime}, \rho^{\prime \prime \prime}(2254))$.
The mass and width for these excited $\rho$ mesons, and the values of the complex parameters
$c_{\omega}$ and $c_i$ in Eq.~(\ref{eq:fpp}) can be found in Ref.~\cite{prd86-032013}.
The explicit expressions of the resonant state functions $GS_\rho, GS_i $, and $BW_\omega$ can be
found, for example, in Ref.~\cite{ly16}.
In this paper, we only consider the contributions from $\rho^{\prime}$ and $\rho^{\prime \prime}$.
Following Ref.~\cite{Wang-2016}, we also assume that
\begin{eqnarray}
F_s(s) = F_t(s) \approx (f_V^{T}/f_V) F_\pi(s)
\end{eqnarray}
for the form factors $F_s(s)$ and $F_t(s)$ that appeared in Eqs.~(\ref{eq:phi2}) and (\ref{eq:phi3}).
In the numerical calculations, we use the Gegenbauer moments
\begin{eqnarray}
a^0_2=0.30\pm 0.05, \quad a^s_2=0.70\pm 0.20, \quad a^t_2=-0.40\pm 0.10,
\end{eqnarray}
for the two-pion distribution amplitudes as used in Ref.~\cite{ly16}.

We here use the same wave functions for the $B$ and $B^0_s$  mesons as those in Refs.~\cite{li2003,Xiao:2011tx}.
The widely used distribution amplitude $\phi_B({\bf k_l})$ is of the form
\begin{eqnarray}
\phi_B(x,b)&=& N_B x^2(1-x)^2\mathrm{exp} \left  [ -\frac{m_B^2\ x^2}{2 \omega_{B}^2} -\frac{1}{2} (\omega_{B}\; b)^2\right] \;.
 \label{phib}
\end{eqnarray}
The normalization factor $N_B$ depends on the value of $\omega_B$ and $f_B$, which is defined through the
normalization relation $\int_0^1dx \; \phi_B(x,b=0)=f_B/(2\sqrt{6})$.
We set $\omega_B = 0.40 \pm0.04$ GeV and $\omega_{B_s}=0.50 \pm 0.05$ GeV~\cite{li2003,Xiao:2011tx}
in the numerical calculations.
The wave function of the final state pseudoscalar meson $P$ ($\pi, K, \eta$, or $\eta^\prime$)
is of the form
\begin{eqnarray}
\Phi_{P}(p_3,x_3)\equiv \frac{i}{\sqrt{2N_C}}\gamma_5
                    \left [ { p \hspace{-2.0truemm}/ }_3 \phi_{P}^{A}(x_3)+m_{03} \phi_{P}^{P}(x_3)
                    + m_{03} ({ n \hspace{-2.2truemm}/ } { v \hspace{-2.2truemm}/ } - 1)\phi_{P}^{T}(x_3)\right ] \;,
\end{eqnarray}
where $m_{03}$ is the chiral mass. The expressions of the relevant distribution amplitudes $\phi_P^{A,P,T}$
can also be found, for example, in Refs.~\cite{ly16,ball99,ball9901,ball05,ball06,prd76-074018,ly14,prd90-114028}.

The mesons $\eta$ and $\eta^\prime$ are considered as the mixtures from $\eta_q$ and $\eta_s$ through the relation
\begin{eqnarray}
\left(\begin{array}{c} \eta \\ \eta^{\prime} \end{array} \right)= \left(\begin{array}{cc}
 \cos{\phi} & -\sin{\phi} \\
 \sin{\phi} & \cos{\phi} \\ \end{array} \right)
 \left(\begin{array}{c}
 \eta_q \\ \eta_s \end{array} \right),
\label{eq:e-ep}
\end{eqnarray}
with the $\eta_q=\left ( u\bar{u}+d\bar{d}\right )/\sqrt{2}$ and $\eta_s=s\bar{s}$.
We adopt the decay constants and mixing angle $\phi$ as~\cite{prd58-114006,plb499-339}
\begin{eqnarray}
f_q=(1.07\pm 0.02)f_{\pi},\quad f_s=(1.34\pm 0.06)f_{\pi},\quad \phi=39.3^\circ\pm 1.0^\circ,
\end{eqnarray}
with $ f_{\pi}=0.131~{\rm GeV}$.

\section{Numerical results}\label{sec:3}  

In numerical calculations, we use the following input parameters (in units of  GeV except $\tau_{B_{s}},\tau_{B^\pm}$)~\cite{pdg2016}:
\begin{eqnarray}
\Lambda^{4}_{ \overline{MS} }&=&0.25, \quad m_{B^{\pm,0}}=5.280, \quad m_{B_s}=5.367,\quad \tau_{B_{s}}=1.510\; ps, \quad \tau_{B^\pm}=1.638\;ps, \nonumber\\
m_{\pi^{\pm}}&=&0.140, \quad m_{\pi^0}=0.135, \quad
m_{K^{\pm}}=0.494, \quad m_{K^0}=0.498,\quad m_{\eta}=0.548, \quad m_{\eta^{\prime}}=0.958.   
\label{eq:inputs}
\end{eqnarray}
The values of the Wolfenstein parameters are the same as those given in Ref.~\cite{pdg2016}:
$A=0.811\pm0.026, \lambda=0.22506\pm 0.00050$, $\bar{\rho} = 0.124^{+0.019}_{-0.018}$, $\bar{\eta}= 0.356\pm 0.011$.

For the decay $B \to P \rho^{\prime}(\rho^{\prime\prime}) \to P \pi \pi$, the differential branching ratio is written
as~\cite{pdg2016}
\begin{eqnarray}
\frac{d{\cal B}}{ds}=\tau_{B}\frac{|\overrightarrow{p_{\pi}}||\overrightarrow{p_P}|}{32\pi^3m^3_{B}}|{\cal A}|^2, \label{expr-br}
\end{eqnarray}
with $\tau_{B}$ the mean lifetime of $B$ meson, and  $s=\omega^2$ the invariant mass squared.
The kinematic variables $|\overrightarrow{p_{\pi}}|$ and $|\overrightarrow{p_P}|$ denote one of the
pion pair's and $P$'s momentum in the center-of-mass frame of the pion pair,
\begin{eqnarray}
 |\overrightarrow{p_{\pi}}|=\frac12\sqrt{s-4m^2_{\pi}}, \quad~~
 |\overrightarrow{p_P}|=\frac12  \sqrt{\big[(m^2_{B}-M_3^2)^2-2(m^2_{B}+M_3^2) s+s^2 \big]/s}. \label{br-momentum}
\end{eqnarray}

\begin{table}[t]
\caption{The PQCD predictions of $\mathcal {B}$ and ${\cal A}_{\rm CP} $ for the quasi-two-body decays
$B_{(s)}\to P (\rho^{\prime} \to) \pi \pi$ and for the decay rates of the two-body decays
$B_{(s)}\to P \rho^{\prime}$. } \label{Presults}
\begin{center}
\begin{tabular}{l c c c}
 \hline \hline
  {Decay modes}       & Quasi-two-body $\mathcal {B}$ (in $10^{-7}$) ~    &~Two-body $\mathcal {B}$  (in $10^{-6}$)   ~
   &${\cal A}_{\rm CP} (\%) $   \\
\hline\hline
  $B^+ \to K^+(\rho^{\prime0}\to)\pi^+ \pi^-$         &$4.66^{+1.05+0.44+0.59+0.11}_{-0.79-0.42-0.50-0.12}$
  &$4.64^{+1.28}_{-1.03}  $
                              &$39^{+5+2+0+0}_{-3-3-1-1} $                 \\
  $B^0 \to K^+(\rho^{\prime-}\to)\pi^- \pi^0$        &$8.88^{+2.66+0.74+0.98+0.26}_{-1.54-0.59-0.84-0.21}$
  &$8.84^{+2.93}_{-1.86}$
                             &$35^{+2+4+0+0}_{-1-4-0-0} $                  \\
  $B_s^0 \to K^-(\rho^{\prime+}\to)\pi^+ \pi^0$      &$13.84^{+5.31+0.04+0.02+0.04}_{-3.59-0.03-0.02-0.04}$
    &$13.78^{+5.29}_{-3.58}$
                               &$25^{+4+1+0+1}_{-4-0-0-1}$                   \\
  $B^+ \to K^0(\rho^{\prime+}\to)\pi^+ \pi^0$        &$10.64^{+2.89+1.63+1.02+0.24}_{-2.16-1.55-0.95-0.22}$
  &$10.60^{+3.47}_{-2.82}$
  &$13^{+3+2+1+0}_{-2-1-0-0}$                  \\

  $B^0 \to K^0(\rho^{\prime0}\to)\pi^+ \pi^-$        &$5.31^{+1.61+0.64+0.41+0.11}_{-1.15-0.60-0.39-0.10}$
  &$5.29^{+1.78}_{-1.35}$
  &$10^{+0+1+1+0}_{-0-0-0-0}$                 \\
  $B_s^0 \to \bar K^0(\rho^{\prime0}\to)\pi^+ \pi^-$  &$0.22^{+0.03+0.01+0.00+0.02}_{-0.02-0.01-0.01-0.02}$
  &$0.22^{+0.04}_{-0.03}$
  &$20^{+15+11+5+0}_{-11-10-5-0}$                 \\
\hline
  $B^+ \to \pi^+(\rho^{\prime0}\to)\pi^+ \pi^-$        &$8.15 ^{+0.00+0.05+1.44+0.22}_{-0.13-0.05-1.30-0.22}$
   &$8.11 ^{+1.45}_{-1.32}$
   &$-29^{+0+2+3+1}_{-0-1-3-0}$                  \\
  $B^0 \to \pi^+(\rho^{\prime-}\to)\pi^- \pi^0$        &$5.32 ^{+0.84+1.58+0.87+0.13}_{-0.79-1.21-0.84-0.14}$
  &$5.30 ^{+1.99}_{-1.67}$
   &$-37^{+0+6+2+1}_{-1-7-2-0}$                 \\
  $B^0 \to \pi^-(\rho^{\prime+}\to)\pi^+ \pi^0$        &$12.34 ^{+4.73+0.66+0.25+0.03}_{-3.30-0.64-0.24-0.02}$
  &$12.29 ^{+4.76}_{-3.36}$
  &$11^{+2+0+0+0}_{-2-1-1-1}$ \\
  $B_s^0 \to \pi^+(\rho^{\prime-}\to)\pi^- \pi^0$      &$0.19^{+0.02+0.07+0.01+0.02}_{-0.01-0.03-0.00-0.01}$
  &$0.19^{+0.08}_{-0.03}$
  &$1^{+0+0+1+2}_{-7-8-1-2}$                  \\
  $B_s^0 \to \pi^-(\rho^{\prime+} \to)\pi^+ \pi^0$     &$0.29 ^{+0.01+0.00+0.02+0.01}_{-0.03-0.01-0.02-0.01}$
  &$0.29 ^{+0.02}_{-0.04}$
  &$-28^{+1+4+1+2}_{-1-8-2-1}$ \\
  $B^+ \to \pi^0(\rho^{\prime+}\to)\pi^+ \pi^0$        &$1.94 ^{+1.50+0.56+0.35+0.00}_{-0.80-0.40-0.29-0.01}$
  &$1.93 ^{+1.63}_{-0.94}$
  &$24^{+2+4+2+2}_{-6-8-2-2}$                  \\
  $B^0 \to \pi^0(\rho^{\prime0}\to)\pi^+ \pi^-$        &$0.26 ^{+0.11+0.04+0.02+0.01}_{-0.08-0.02-0.02-0.01}$
  &$0.26 ^{+0.12}_{-0.09}$
  &$-54^{+5+7+5+0}_{-4-4-5-1}$ \\
  $B_s^0 \to \pi^0(\rho^{\prime0} \to)\pi^+ \pi^-$     &$0.15 ^{+0.03+0.02+0.00+0.01}_{-0.02-0.01-0.00-0.01}$
  &$0.15 ^{+0.04}_{-0.02}$
  &$-30^{+0+10+7+2}_{-8-15-0-0}$ \\
\hline
  $B^+ \to \eta(\rho^{\prime+}\to)\pi^+ \pi^0$         &$4.41^{+1.55+0.29+0.09+0.01}_{-1.08-0.27-0.09-0.00}$
   &$4.39^{+1.57}_{-1.11}$
   &$2^{+1+0+0+0}_{-2-0-1-0}$              \\
  $B^0 \to \eta(\rho^{\prime0}\to)\pi^+ \pi^-$         &$0.14 ^{+0.02+0.02+0.01+0.00}_{-0.02-0.02-0.01-0.01}$
  &$0.14 ^{+0.02}_{-0.03}$
  &$-13 ^{+1+6+1+2}_{-3-3-1-2}$                 \\
  $B_s^0 \to \eta(\rho^{\prime0}\to)\pi^+ \pi^-$       &$0.08 ^{+0.03+0.00+0.00+0.00}_{-0.02-0.00-0.00-0.00}$
  &$0.08 ^{+0.03}_{-0.02}$
  &$37^{+0+0+0+1}_{-1-1-1-1}$                 \\
  $B^+ \to \eta^{\prime}(\rho^{\prime+}\to)\pi^+ \pi^0$ &$3.21^{+1.09+0.17+0.02+0.00}_{-0.77-0.15-0.02-0.01}$
  &$3.20^{+1.10}_{-0.78}$
  &$46^{+5+3+2+0}_{-3-2-1-0} $                 \\
  $B^0 \to \eta^{\prime}(\rho^{\prime0}\to)\pi^+ \pi^-$  &$0.22 ^{+0.07+0.02+0.01+0.01}_{-0.04-0.00-0.01-0.00}$
  &$0.22 ^{+0.07}_{-0.04}$
  &$20^{+10+27+8+1}_{-11-30-7-0}$                 \\
  $B_s^0 \to \eta^{\prime}(\rho^{\prime0}\to)\pi^+ \pi^-$ &$0.17 ^{+0.06+0.00+0.00+0.00}_{-0.05-0.01-0.00-0.00}$
  &$0.17 ^{+0.06}_{-0.05}$
  &$54^{+1+1+1+0}_{-0-0-0-0}$                 \\
 \hline \hline
\end{tabular}
\end{center}
\end{table}
\begin{table}[t]
\caption{The PQCD predictions of $\mathcal {B}$ and ${\cal A}_{\rm CP} $ for the quasi-two-body decays
$B_{(s)}\to P (\rho^{\prime\prime} \to) \pi \pi$ and for the decay rates of the two-body decays
$B_{(s)}\to P \rho^{\prime\prime}$. }
\label{Vresults}
\begin{center}
\begin{tabular}{l c c c}
 \hline \hline
  {Decay modes}       & Quasi-two-body $\mathcal {B}$ (in $10^{-7}$) ~    &~Two-body $\mathcal {B}$  (in $10^{-6}$)   ~
   &$\cal A_{\rm CP} (\%)$   \\
\hline\hline
  $B^+ \to K^+(\rho^{\prime\prime0}\to)\pi^+ \pi^-$    &$2.53^{+0.69+0.29+0.35+0.07}_{-0.52-0.27-0.31-0.05}$
  &$3.12 ^{+1.02}_{-0.82}$
  &$33^{+6+1+0+0}_{-7-1-0-1}$                  \\
  $B^0 \to K^+(\rho^{\prime\prime-}\to)\pi^- \pi^0$              &$4.80^{+1.51+0.51+0.56+0.12}_{-1.08-0.40-0.52-0.09}$
  &$5.92^{+2.09}_{-1.56}$
  &$29^{+2+5+1+1}_{-5-4-1-0}$                  \\
  $B_s^0 \to K^-(\rho^{\prime\prime+}\to)\pi^+ \pi^0$            &$6.52^{+2.49+0.02+0.01+0.02}_{-1.69-0.02-0.01-0.01}$
  &$8.03^{+3.07}_{-2.08}$
  &$26^{+4+0+0+1}_{-4-1-0-1}$                   \\
  $B^+ \to K^0(\rho^{\prime\prime+}\to)\pi^+ \pi^0$              &$6.20^{+1.90+1.04+0.73+0.13}_{-1.43-0.95-0.65-0.11}$
  &$7.64 ^{+2.82}_{-2.27}$
  &$14^{+3+1+0+0}_{-4-3-1-1}$                  \\

  $B^0 \to K^0(\rho^{\prime\prime0}\to)\pi^+ \pi^-$              &$2.98^{+1.02+0.40+0.28+0.07}_{-0.74-0.37-0.27-0.06}$
  &$3.67^{+1.40}_{-1.08}$
  &$9^{+1+1+1+1}_{-0-0-0-0}$                 \\
  $B_s^0 \to \bar K^0(\rho^{\prime\prime0}\to)\pi^+ \pi^-$       &$0.11^{+0.01+0.01+0.00+0.01}_{-0.01-0.00-0.00-0.01}$
  &$0.14 \pm0.02$
  &$1^{+16+13+6+1}_{-10-11-5-0}$                 \\
\hline
  $B^+ \to \pi^+(\rho^{\prime\prime0}\to)\pi^+ \pi^-$           &$2.81 ^{+0.28+0.02+0.56+0.09}_{-0.40-0.03-0.52-0.09}$
  &$3.46 ^{+0.78}_{-0.82}$
  &$-35^{+3+1+2+0}_{-1-2-5-1}$                  \\
  $B^0 \to \pi^+(\rho^{\prime\prime-}\to)\pi^- \pi^0$           &$1.28 ^{+0.13+0.41+0.25+0.03}_{-0.09-0.11-0.17-0.03}$
  &$1.58^{+0.61}_{-0.28}$
  &$-51^{+1+8+2+1}_{-2-0-2-1}$                 \\
  $B^0 \to \pi^-(\rho^{\prime\prime+}\to)\pi^+ \pi^0$          &$5.61 ^{+2.16+0.38+0.14+0.02}_{-1.50-0.34-0.13-0.01}$
  &$6.92^{+2.71}_{-1.90}$
  &$11^{+3+1+1+1}_{-2-0-0-0}$ \\
  $B_s^0 \to \pi^+(\rho^{\prime\prime-}\to)\pi^- \pi^0$        &$0.08^{+0.01+0.04+0.00+0.01}_{-0.01-0.02-0.00-0.01}$
  &$0.10 ^{+0.05}_{-0.03}$
  &$-3^{+5+8+3+2}_{-1-3-3-0}$                  \\
  $B_s^0 \to \pi^-(\rho^{\prime\prime+} \to)\pi^+ \pi^0$       &$0.16 ^{+0.02+0.01+0.02+0.01}_{-0.01-0.01-0.01-0.00}$
  &$0.20 ^{+0.04}_{-0.02}$
  &$-24^{+2+8+2+2}_{-0-8-0-1}$ \\
  $B^+ \to \pi^0(\rho^{\prime\prime+}\to)\pi^+ \pi^0$          &$0.67 ^{+0.60+0.27+0.15+0.00}_{-0.29-0.15-0.13-0.00}$
  &$0.83^{+0.83}_{-0.43}$
  &$18^{+0+7+5+2}_{-4-18-7-2}$                  \\
  $B^0 \to \pi^0(\rho^{\prime\prime0}\to)\pi^+ \pi^-$          &$0.14 ^{+0.04+0.03+0.01+0.00}_{-0.03-0.02-0.01-0.00}$
  &$0.17^{+0.06}_{-0.05}$
  &$-53^{+0+8+4+0}_{-2-3-6-1}$ \\
  $B_s^0 \to \pi^0(\rho^{\prime\prime0} \to)\pi^+ \pi^-$       &$0.06 ^{+0.01+0.02+0.00+0.01}_{-0.01-0.01-0.00-0.00}$
  &$0.07 ^{+0.03}_{-0.02}$
  &$-35^{+0+19+1+2}_{-2-14-0-2}$ \\
\hline
  $B^+ \to \eta(\rho^{\prime\prime+}\to)\pi^+ \pi^0$           &$2.11^{+0.77+0.17+0.05+0.00}_{-0.53-0.14-0.04-0.00}$
  &$2.60 ^{+0.97}_{-0.68}$
  &$2^{+1+0+0+0}_{-1-0-0-0}$              \\
  $B^0 \to \eta(\rho^{\prime\prime0}\to)\pi^+ \pi^-$           &$0.08 ^{+0.02+0.02+0.00+0.00}_{-0.02-0.01-0.00-0.00}$
  &$0.10\pm 0.03$
  &$-32 ^{+4+3+5+1}_{-1-0-5-0}$                 \\
  $B_s^0 \to \eta(\rho^{\prime\prime0}\to)\pi^+ \pi^-$         &$0.04 ^{+0.01+0.00+0.00+0.00}_{-0.01-0.00-0.00-0.00}$
  &$0.05\pm0.01$
  &$44^{+1+1+0+0}_{-2-1-1-1}$                 \\
  $B^+ \to \eta^{\prime}(\rho^{\prime\prime+}\to)\pi^+ \pi^0$  &$1.49^{+0.52+0.09+0.01+0.01}_{-0.38-0.08-0.01-0.00}$
  &$1.84 ^{+0.65}_{-0.48}$
  &$50^{+0+3+2+0}_{-1-3-2-0}$                 \\
  $B^0 \to \eta^{\prime}(\rho^{\prime\prime0}\to)\pi^+ \pi^-$  &$0.11 ^{+0.02+0.01+0.00+0.00}_{-0.02-0.01-0.01-0.00}$
  &$0.14\pm0.03$
  &$31^{+10+29+8+1}_{-12-34-7-0}$                  \\
  $B_s^0 \to \eta^{\prime}(\rho^{\prime\prime0}\to)\pi^+ \pi^-$ &$0.08 ^{+0.02+0.00+0.00+0.00}_{-0.02-0.00-0.00-0.00}$
  &$0.10 \pm0.02$
  &$60^{+1+0+0+0}_{-1-0-0-0}$                 \\
 \hline \hline
\end{tabular}
\end{center}
\end{table}

By using the differential branching fraction, Eq.~(\ref{expr-br}), and the decay amplitudes
in the Appendix of Ref.~\cite{ly16},
we calculate the $CP$ averaged branching ratios ($\cal B$) and direct $CP$-violating
asymmetries ($\cal A_{CP}$) for the decays $B_{(s)}\to P (\rho^{\prime} \to) \pi \pi $ and list the
results in Table~\ref{Presults}.
Meanwhile, $\cal B$ and $\cal A_{CP}$ for the decays $B_{(s)}\to P (\rho^{\prime\prime} \to )\pi \pi $ are
shown in Table~\ref{Vresults}.
The four errors of these PQCD predictions as listed in Tables~\ref{Presults} and~\ref{Vresults}
come from the uncertainties of $\omega_B/\omega_{B_s}$, $a^t_2=-0.40 \pm 0.10$,
$a^s_2=0.70 \pm 0.20$, and $a^0_2=0.30 \pm 0.05$, respectively.

For $B^+\to K^+(\rho^{\prime 0}\to)\pi^+\pi^-$ and the other three $B\to K (\rho^\prime \to)\pi\pi$
decay modes, the PQCD predictions for their branching ratios as listed in Table I are a
little different from those as given previously in Table I of Ref.~[16]. The reason
is very simple:  the Gegenbauer moments $a_2^{t,s,0}$  used here have been  modified slightly
from those in Ref.~[16] as discussed in Ref.~[18].

\begin{figure}[tbp]
\vspace{-0.5 cm}
\centerline{\epsfxsize=7.5cm \epsffile{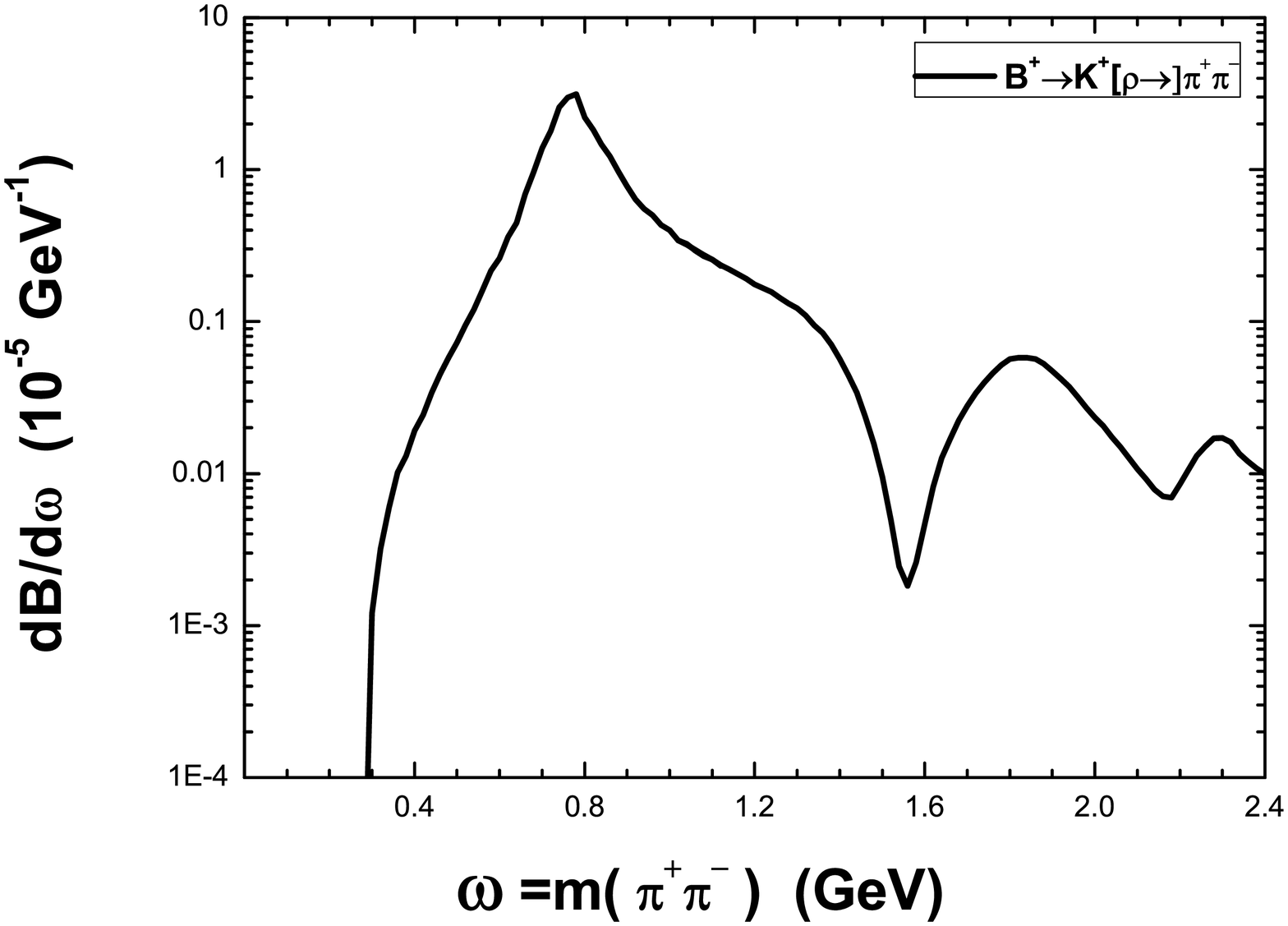}
            \epsfxsize=7.5cm \epsffile{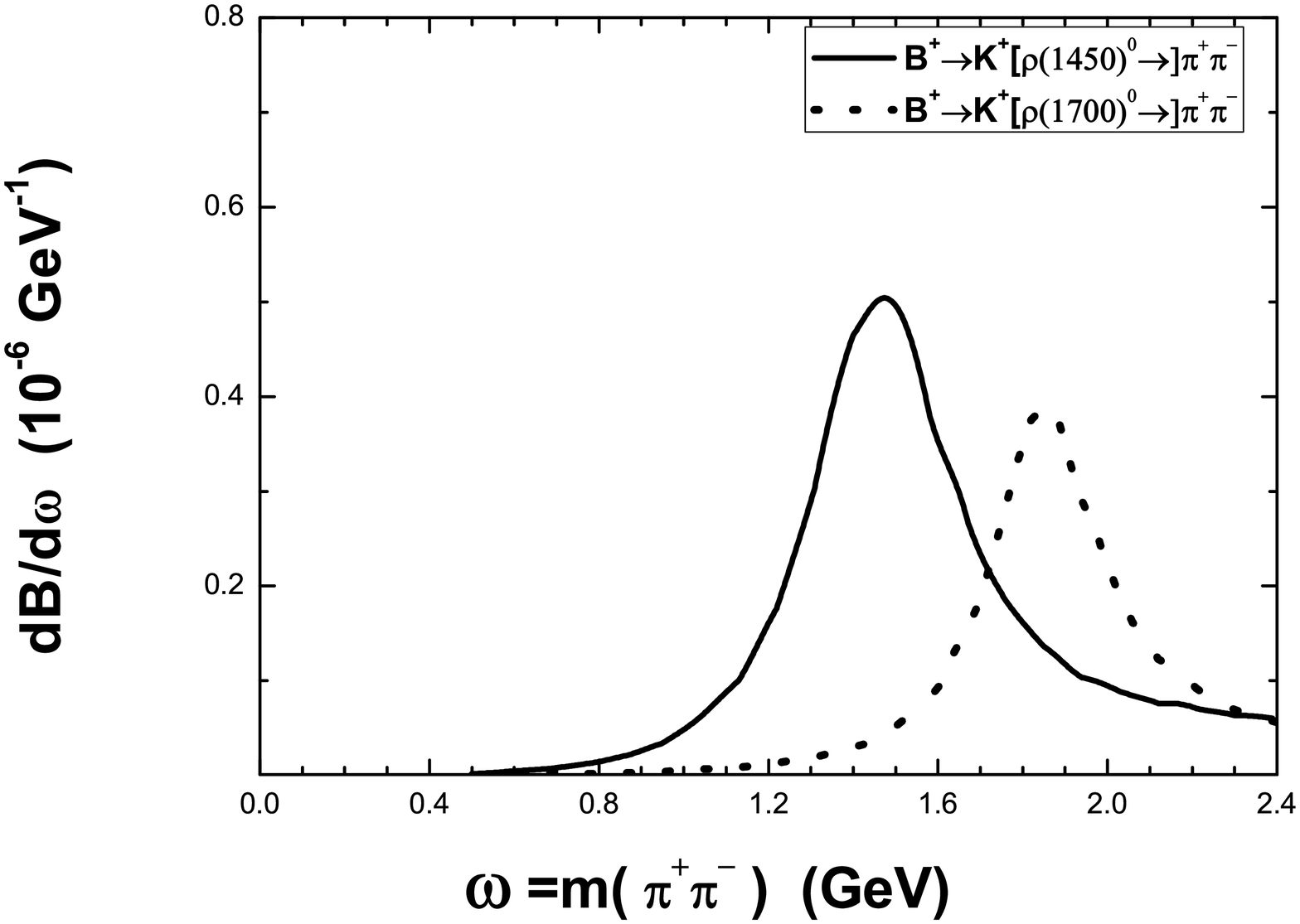}}
\vspace{-0.2cm}
  {\scriptsize\bf (a)\hspace{7.7cm}(b)}
\caption{(a) The summation of the contributions from $\rho(770), \rho^\prime$, and $ \rho^{\prime\prime}$ for the
                   differential branching ratios of the $B^+\to K^+\rho \to K^+\pi^+\pi^-$ decays.
        (b)The comparison of the differential branching distributions for $B^+\to K^+\rho^{\prime0}\to K^+\pi^+\pi^-$
                   (solid curve) and $B^+ \to K^+\rho^{\prime\prime0}\to K^+\pi^+\pi^-$ (dashed curve).}
\label{fig-dis-br}
\end{figure}

Taking the quasi-two-body decay $B^+ \to \pi^+\rho^{\prime0} \to \pi^+\pi^+ \pi^-$ as an example,
the PQCD prediction for its branching ratio  and $CP$-violating asymmetry ${\cal A}_{\rm CP}$ are the following:
\begin{eqnarray}
\mathcal{B}( B^+ \to \pi^+(\rho^{\prime 0} \to ) \pi^+ \pi^-)&=& \left (8.15^{+1.46}_{-1.33} \right )\times 10^{-7}, \label{eq:br01}\\
{\cal A}_{\rm CP}(B^+ \to \pi^+(\rho^{\prime 0} \to ) \pi^+ \pi^-)&=& \left (-29^{+4}_{-3} \right )\%. \label{eq:acp01}
\end{eqnarray}
Here the individual errors as listed in Table~\ref{Presults} have been added in quadrature.
Such a PQCD prediction for its branching ratio agrees well with the measured value
$(1.4^{+0.6}_{-0.9})\times10^{-6}$ from ${\it BABAR}$ Collaboration within errors~\cite{prd79-072006}.
Furthermore, the PQCD prediction $\mathcal{A_{\rm CP}}=(-29^{+4}_{-3})\% $ for this decay mode
is also consistent with the measured value $(-6 \pm 28\pm 20^{+12}_{-35})\%$ from ${\it BABAR}$~\cite{prd79-072006}.

The width $\Gamma_{\pi\pi}$ for the $\rho^\prime\to \pi\pi$ process was found to be $\sim 22$~MeV
in Ref.~\cite{ijmpa13-5443}, which is consistent with the value $17\sim 25$~MeV estimated from the $e^+e^-$
annihilation experiments~\cite{zpc62-455}.
The branching fraction ${\mathcal B}(\rho^\prime\to\pi\pi)=(4.6-10)\% $ could be induced with the
$\Gamma_{\rho^\prime}=0.311\pm0.062$ GeV~\cite{zpc62-455}.
The $\rho^\prime\to \pi\pi$ branching fraction, on the other hand, could be estimated from the
relation~\cite{EPJC2-269}
\begin{eqnarray}
\Gamma_{\rho^\prime\to\pi\pi}=\frac{g^2_{\rho^\prime\pi\pi}}{6\pi}
\frac{|\overrightarrow{p_\pi}(m^2_{\rho^\prime})|^3}{m^2_{\rho^\prime}},
\label{bf}
\end{eqnarray}
where the coupling $g_{\rho^\prime\pi\pi}$ is fetched from the $\rho^\prime$ component of the timelike form factor
$F_\pi$ in Eq.~(\ref{eq:fpp}) according to $F^{\rho^\prime}_\pi(\omega^2)\approx g_{\rho^\prime\pi\pi} \omega f_{\rho^\prime}/D_{\rho^\prime}(\omega^2)$
at $\omega=m_{\rho^\prime}$.
The decay constant $f_{\rho^\prime}=0.185^{+0.030}_{-0.035}$~GeV
resulting from the data $\Gamma_{\rho^\prime\to e^+e^-}=1.6-3.4$
keV~\cite{zpc62-455} is adopted in this work, which agrees with
$f_{\rho^\prime}=(0.186\pm0.014)$~GeV from the double-pole QCD sum rules~\cite{1205-6793},
$f_{\rho^\prime}=(0.182\pm0.005)$~GeV from the perturbative analysis in the large-$N_c$ limit~\cite{prd77-116009},
or $f_{\rho^\prime}=0.128$~GeV from the relativistic constituent quark model~\cite{prd60-094020}.
Utilizing Eq.~(\ref{bf}), we find  ${\mathcal B}(\rho^\prime\to\pi\pi)=10.04^{+5.23}_{-2.61}\%$.
From the definition of the decay rates between the quasi-two-body and the corresponding two-body decay modes
\begin{eqnarray}
\mathcal{B}( B_{(s)} \to P (\rho^{\prime } \to ) \pi \pi ) =
\mathcal{B}( B_{(s)} \to P \rho^{\prime } ) \cdot {\mathcal B}(\rho^\prime \to\pi\pi), \label{eq:def1}
\end{eqnarray}
we then can find the PQCD predictions for  $ {\cal B}( B/B_s \to P\rho^{\prime})$, as listed
in the third column of Table~\ref{Presults}, where the individual errors have been added in quadrature.

For the cases of the considered quasi-two-body and two-body decays involving $\rho^{\prime\prime}$ instead of
$\rho^\prime$, in principle, one can obtain the PQCD predictions for the branching ratios and $CP$-violating
asymmetries in a similar way as the case for $\rho^\prime$.
But, there is not much reliable information about the properties of the $\rho^{\prime\prime}$ meson except
its mass and width ($m_{\rho^{\prime\prime}}=1.72\pm 0.02$ GeV and $\Gamma_{\rho^{\prime\prime}} =0.25\pm 0.10$ GeV)~\cite{pdg2016}.
What we can do here is to make some rough estimations of the branching ratios and $CP$ violating asymmetries for
the considered $B \to P\rho^{\prime\prime}$ decays, and list the PQCD predictions in Table~\ref{Vresults}.
For given $\Gamma_{\rho^{\prime\prime} \to e^+e^-}=0.69\pm0.15$~keV ~\cite{zpc62-455}, we find
the longitudinal decay constant $f_{\rho^{\prime\prime}}=0.103^{+0.011}_{-0.012}$~GeV.
And then ${\mathcal B}(\rho^{\prime\prime} \to \pi\pi)=8.11^{+2.22}_{-1.47}\%$ can be obtained by using the same
methods as for the decays involving $\rho^{\prime}$.
The errors of the PQCD predictions listed in the third column of Table~\ref{Vresults} have been added in
quadrature.

In Fig.~\ref{fig-dis-br}(a), we show the $\omega$ dependence of the differential decay rate
$d{\cal B}(B^+ \to K^+ \pi^+\pi^-)/d\omega$ after the inclusion of the contributions from the resonant state
$\rho(770)$, $\rho^{\prime}$, and $\rho^{\prime\prime}$.
One can see that there exists a clear dip near $\omega=m(\pi^+\pi^-) \sim 1.6~{\rm GeV}$ in Fig.~\ref{fig-dis-br}(a).
The position of this dip and the pattern of the whole curve do agree well with Fig.~45 of 
Ref.~\cite{prd86-032013}, where the pion form factor$-$squared  $|F_\pi|^2$
measured by ${\it BABAR}$ are illustrated as a function of $\sqrt{s^\prime}$ [i.e., $m(\pi\pi)$]
in the region from $0.3$ to $3$ GeV.
In fact, the differential decay rate  $d{\cal B}/d\omega$ does depend on the
values of $|F_\pi|^2$.
In Fig.~\ref{fig-dis-br}(a), we find the prominent $\rho(770)$ peak, a shoulder around the $\rho^{\prime}(1450)$ and
a clear dip followed by an enhancement (second a little lower and wide peak) in the $\rho^{\prime\prime}(1700)$
region.
The clear dip at $\omega \approx 1.6~{\rm GeV}$ is caused by the destructive interference
between the resonant state $\rho^{\prime}$ and $\rho^{\prime\prime}$.
We calculated numerically the interference terms between $\rho^{\prime}$ and $\rho^{\prime\prime}$ amplitudes and
found the large negative contribution to the branching ratios.

Taking $B^+ \to K^+\rho^{\prime 0} \to K^+\pi^+\pi^- $ and $B^+ \to K^+ \rho^{\prime\prime 0} \to K^+\pi^+\pi^- $
decays as examples, we found the PQCD predictions for the individual decay rate and the interference term,
\begin{eqnarray}
{\cal B}(B^+ \to K^+(\rho^{\prime 0}\to ) \pi^+\pi^-) &=& 4.66^{+1.10}_{-1.05} \times 10^{-7}, \nonumber\\
{\cal B}(B^+ \to K^+(\rho^{\prime \prime 0}\to ) \pi^+\pi^-) &=& 2.53^{+0.90}_{-0.67} \times 10^{-7}, \nonumber\\
{\rm interference \ \  term} &\approx & -4.55 \times 10^{-7}. \label{eq:inter}
\end{eqnarray}
One can see that the interference term is indeed large and negative when compared with the other two individual
contributions, which in turn results in a clear dip in the region around $\omega \approx 1.6$ GeV.

In Fig.~\ref{fig-dis-br}(b), we show the PQCD prediction for the $\omega$ dependence of the differential decay rate
$d{\cal B}(B^+ \to K^+ \pi^+\pi^-)/d\omega$, when the contribution from the resonance
$\rho^{\prime}$ (solid line) and $\rho^{\prime\prime}$ (dashed line) is taken into account, respectively.
The decay rate for the $B^+ \to K^+(\rho^{\prime0}\to) \pi^+\pi^-$ decay is a little larger
than that for the $B^+ \to K^+(\rho^{\prime\prime0}\to) \pi^+\pi^-$ decay.
The difference is mainly governed by the different parameters $c_{\rho^\prime}$ and $c_{\rho^{\prime\prime}}$,
as well as the parameters of the corresponding $GS_{\rho^\prime}$ and $GS_{\rho^{\prime\prime}}$ function
in Eq.~(\ref{eq:fpp}).

\section{Summary}

In this work, we calculated the quasi-two-body decays
$B \to  P \rho^{\prime}(1450) \to  P \pi\pi$ and $B \to  P \rho^{\prime\prime}(1700)\to  P \pi\pi$
with $ P =(\pi,K,\eta,\eta^\prime )$ by utilizing the vector current timelike form factor $F_\pi(s)$
with the inclusion of the final state interactions between pion pairs in the resonant regions
associated with $\rho^\prime(1450)$ and $\rho^{\prime\prime}(1700)$.
\begin{enumerate}
\item
The PQCD predictions for the $CP$-averaged branching ratios and direct $CP$-violating asymmetries of the
considered quasi-two-body decays have been listed in Tables I and II.
The decay rates for the considered decay modes are generally in the order of $10^{-7}$ to $ 10^{-9}$.

\item
The whole pattern of the pion form factor$-$squared  $|F_\pi|^2$ measured by ${\it BABAR}$ Collaboration could be
understood based on our studies, as illustrated by Fig.~1(a): where one can see the prominent $\rho(770)$ peak,
a shoulder around $\rho^{\prime}(1450)$, a clear dip at $\omega \approx 1.6~{\rm GeV}$ caused by the
destructive interference between the contributions from $\rho^{\prime}(1450)$ and $\rho^{\prime\prime}(1700)$,
and an enhancement in the $\rho^{\prime\prime}(1700)$ region.

\item
The branching ratios of the corresponding two-body decays have been extracted from the quasi-two-body decay modes.
More precise data from the LHCb and the future Belle II will test our predictions.

\end{enumerate}

\begin{acknowledgments}
Many thanks to Hsiang-nan Li and Rui Zhou for valuable discussions.
This work was supported by the National Natural Science Foundation of China under the Grants No.~11235005
and No.~11547038.

\end{acknowledgments}


\end{document}